\documentclass[twocolumn,amsmath,floats,showpacs,nofootinbib]{revtex4}
\usepackage[usenames]{color}
\usepackage{graphicx}
\usepackage{epstopdf}
\usepackage{textcomp}
\usepackage{hyperref}
\usepackage{multirow}
\usepackage{tikz}
\usepackage{rotating}
\hypersetup{colorlinks=true}

\begin{document}

\title{Structure and physical properties of superconducting compounds Y(La)-Ba(Sr)-Cu-O}

\author{B.I. Verkin, B.B. Banduryan, A.S. Barylnik, A.G. Batrak, N.L. Bobrov, I.S. Braude, Yu.L. Gal'chinetskaya, A.L. Gaiduk, A.M. Gurevich, V.V. Demirskii, V.I. Dotsenko, V.I. Eropkin, S.V. Zherlitsyn, A.P. Isakina, I.F. Kislyak, V.A. Konovodchenko, F.F. Lavrent'ev, L.S. Litinskaya, V.A. Mikheev, V.I. Momot, V.D. Natsik,
I.N. Nechiporenko, A.S. Panfilov, Yu.A. Pokhil, A.I. Prokhvatilov, E.Ya. Rudavskii, L.F. Rybal'chenko, I.V. Svechkarev, A.M. Stepanenko, M.A. Strzhemechnyi, L.I. Fedorchenko,V.D. Fil', V.V. Fisun, V.G. Khomenko, V.K. Chagovets, G.A. Sheshin, I.K. Yanson, and V.V. Sergienko}
\affiliation{B.Verkin Institute for Low Temperature Physics and Engineering, 47, Lenin Ave., 310164 Kharkov, Ukraine
Email address: bobrov@ilt.kharkov.ua}
\published {\href{http://fntr.ilt.kharkov.ua/fnt/pdf/13/13-7/f13-0771r.pdf}{Fiz. Nizk. Temp.}, \textbf{13}, 771 (1987)); (Sov. J. Low Temp. Phys., \textbf{13}, 442(1987)}
\date{\today}

\begin{abstract}The structure and physical properties of superconducting compounds Y(La)-Ba(Sr)-Cu-O are studied, the compounds being prepared by the method of cryogenic dispersion of a charge consisting of premix oxides and carbonates. Electrical conductivity and critical current density of the superconductors are measured over a wide temperature range of 10~$mK$ to 300~$K$. Degradation of the superconductor critical parameters in time and structural characteristics, magnetic susceptibility, heat capacity and acoustic properties are studied, and current-voltage characteristics are determined.


\end{abstract}

\maketitle

The purpose or the present article is to present a study of the structure and physical properties of the high- temperature superconductors Y(La)-Ba(Sr)-Cu-O, prepared by known technology \cite{1,2,3}. What was essentially new in the technological process of obtaining superconducting ceramics was the use of cryogenic dispersion of a charge of premix oxides and carbonates by a method described in Ref.\cite{4}. Particles of premix oxides and salts with initial dimension of the order of $10 \mu m$ after dispersion had dimensions of $1 \mu m$ or less. From the cryopowder after the passage of the solid phase reaction, mechanically durable samples of given thickness (from 0.1 to 3 mm) and variable diameter were obtained. During electrical measurements copper contacts (glued on with the help of silver paste) and platinum contacts of diameter $30-70 \mu m$ were used for electrodes.
\section{SURFACE MORPHOLOGY OF THE SAMPLES}
Sample surfaces were studied by means of a scanning electron microscope (model BS-300, accelerating voltage 25~$kV$, regime of secondary electrons, without dusting of the conducting layer); the range of magnification was 50 to $40,000\times$.
\begin{figure*}[]
\includegraphics[width=16cm,angle=0]{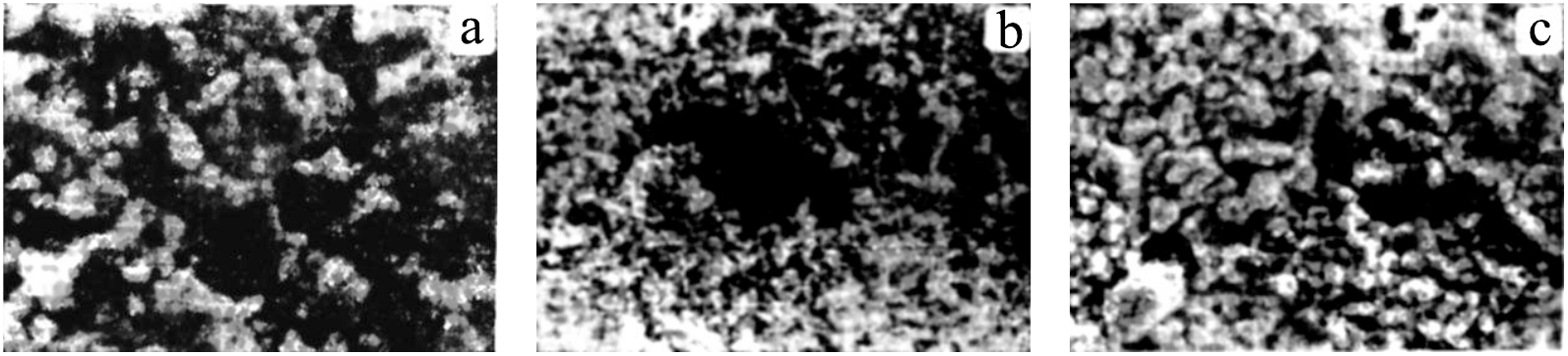}
\caption[]{Surface morphology of the compounds: a) isolated particles and aggregates ($\times 500$) in $\rm La_{2-x}Sr_xCuO_4$; b) micropores and macropores ($\times 1250$) in $\rm La_{2-x}Sr_xCuO_4$; c) isolated particles in $\rm Y_{1.2}Ba_{0.8}CuO_4$.}
\label{Fig1}
\end{figure*}

\begin{figure}[]
\includegraphics[width=8cm,angle=0]{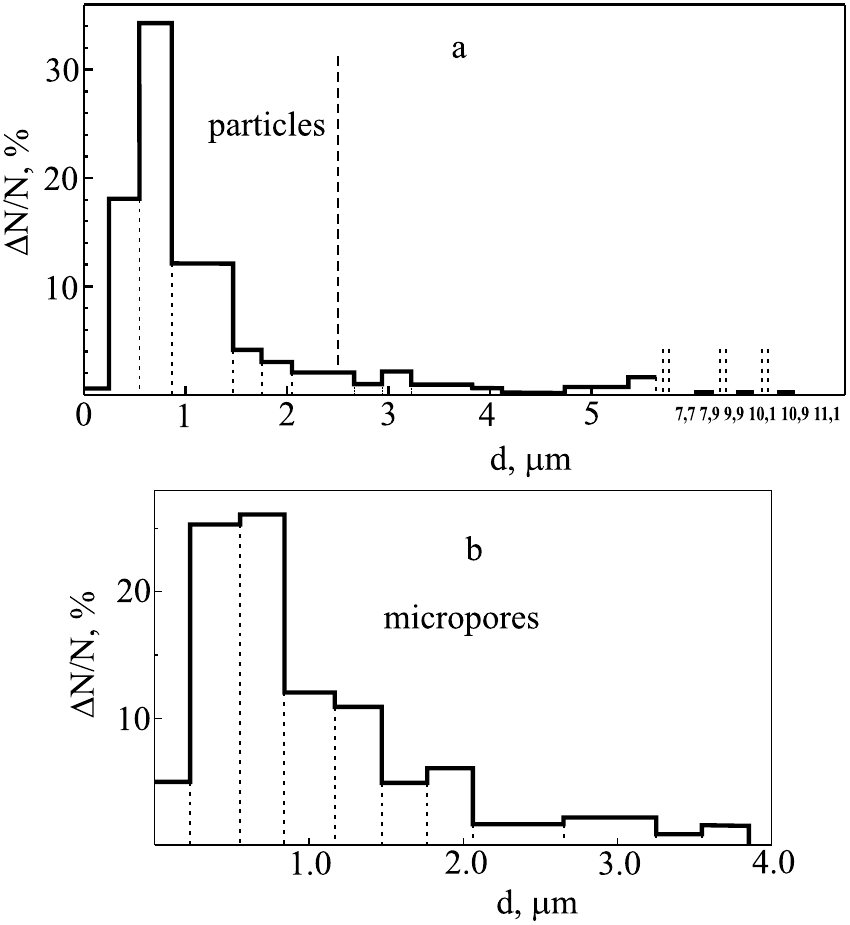}
\caption[]{Histograms of the size distributed of a) particles and aggregates and b) micropores in $\rm La_{2-x}Sr_xCuO_4$.}
\label{Fig2}
\end{figure}
The characteristic elements of the surface of the ceramic appear to be particles of spherical shape and pores of irregular shape. The spherical particles are, as a rule, distributed individually, with aggregates of a few particles appearing occasionally. The diameters of the particles fluctuate within the limits 0.2-2~$\mu m$, while the sizes of the aggregates cover the range 23-10~$\mu m$. Under observation it is possible to divide the pores into groups by size, micropores sized 0.2-3~$\mu m$, and macro-pores having diameters in the range 20-30~$\mu m$. These latter are met rather rarely and are probably caused by insufficient pressure during the pressing of the samples, nonconglomeration of particles or gas emission during conglomeration. Typical photographs of sample surfaces are shown in Figs \ref{Fig1}a and \ref{Fig1}b. The surface structure of a sample of YBaCuO is shown in Fig.\ref{Fig1}c. Results of statistical analysis of the microphotographs for $\rm La_{1.8}Sr_{0.2}CuO_4$ are presented in Fig.\ref{Fig2}. It can be seen that the distribution function of the micropores as well as that of the particles has a sharply defined maximum at a characteristic dimension 0.6-0.8~$\mu m$, which testifies to a sufficiently homogeneous dispersion of the premix powder.
\section{X-RAY STRUCTURE STUDIES OF SUPERCONDUCTORS IN THE RANGE 4.5-293 K}

Measurements were made with a DRON-0.5 x-ray diffractometer with $K_{\alpha}$-irradiation from a copper anode at temperatures in the range 4.5-293~$K$. The samples were cooled in a special helium cryostat for x-ray studies. Qualitative photograms and diffractograms were obtained, the analysis of which has allowed us to conclude that the samples are 95-97\% in the ground phase. The dimension of the region of coherent scattering in the samples under study stood in the range $10^{-4}-10^{-5}\ cm$. Strains were essentially absent.

\begin{figure}[]
\includegraphics[width=8.5cm,angle=0]{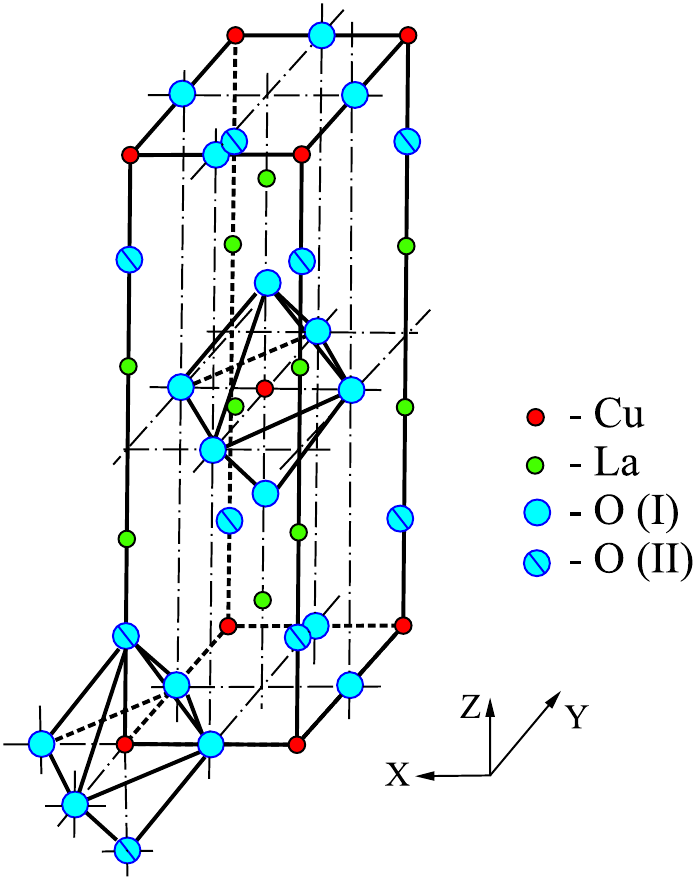}
\caption[]{Crystalline structure of $La_{2-x}Sr_xCuO_4$ at $T=293~K$ (space group $I4/mmm$), z=2.}
\label{Fig3}
\end{figure}

From indexing the x-ray patterns of the compound $\rm La_{2-x}Sr_xCuO_4$ it was established that the observed reflections emanated from crystals of tetragonal symmetry with lattice parameters $a=3.761_6\ \text{\AA}$ and $c=13.174_5\ \text{\AA}$ at 293~$K$, which are somewhat smaller than the values in the literature \cite{3}. From the observed fading out of the reflections it was also established that the tetragonal lattice of this compound is volume-centered, which further advances the assumption of isostructurality of the ceramic under study with its analog $\rm La_{2-x}Ba_xCuO_4$, for which the full symmetry of the lattice has already been determined \cite{5}. On the basis of this assumption, we modeled the arrangement of the atoms of two formula units in the unit cell with the help of the program "Polikristall-3" on an ES-1033 computer and calculated the intensities of the x-ray reflections, taking into account all scattering factors. From this it was established that at room temperature the compound $\rm La_{2-x}Sr_xCuO_4$ has the crystal structure of a layered perovskite of the spatial symmetry group $I4/mmm$. The structure of $\rm La_{2-x}Sr_xCuO_4$ is schematically
represented in Fig.\ref{Fig3}, and the experimental and calculated values of the angles and intensities of reflections for part of the diffractogram are given in Table \ref{tab1}. The coordinates of the atoms, the symmetry of their pointwise positions and the coefficients $B$ of the isotropic thermal factor are given in Table \ref{tab2}.
\begin{table}[]
\caption[]{Experimental and Calculated Angles ($2\theta$) and Relative Intensities ($I$) of Reflections ($hkl$) in $\rm La_{1.8}Sr_{0.2}CuO_4$ at 293~$K$.}
\begin{tabular}{|c|c|c|c|c|} \hline
\multirow{2}{*}\ \ \ {$hkl$}\ \ \  & \multicolumn{2}{|c|}{Experiment} & \multicolumn{2}{|c|}{Calculated} \\ \cline{2-5}
 & $2\theta_{ex}$, deg & \ \ \ $I_{ex}$\ \ \  & $2\theta_{ca}$, deg & \ \ \ $I_{ca}$\ \ \  \\ \hline
002 & --- & --- & 13.44 & 34 \\ \hline
011 & 24.63 & 230 & 24.61 & 213 \\ \hline
004 & 27.08 & 109 & 27.07 & 120 \\ \hline
013 & 31.32 & 1000 & 31.31 & 1000 \\ \hline
110 & 33.70 & 640 & 33.70 & 631 \\ \hline
112 & 36.50 & 31 & 36.43 & 12 \\ \hline
006 & 41.10 & 98 & 41.11 & 104 \\ \hline
015 & 41.85 & 167 & 41.86 & 186 \\ \hline
114 & 43.76 & 225 & 43.75 & 238 \\ \hline
020 & 48.43 & 240 & 48.40 & 269 \\ \hline
022 & --- & --- & 50.47 & 2 \\ \hline
116 & 54.19 & 84 & 54.17 & 91 \\ \hline
017 & 54.56 & 84 & 54.52 & 96 \\ \hline
121 & 55.06 & 37 & 55.04 & 40 \\ \hline
008 & 55.78 & 33 & 55.82 & 37 \\ \hline
024 & 56.41 & 48 & 56.34 & 53 \\ \hline
123 & 58.78 & 230 & 58.78 & 254 \\ \hline

\end{tabular}
\label{tab1}
\end{table}

\begin{table}[]
\caption[]{Coordinates, Pointwise Symmetry and Coefficients of the Thermal Factor of Independent Atoms in the Unit Cell of $\rm La_{1.8}Sr_{0.2}CuO_4$ (Spatial Group $I4/mmm$)}
\begin{tabular}{|c|c|c|c|c|c|c|} \hline
Type of & Multiplicity &Pointwise&\multicolumn{3}{|c|}{Coordinates} & \ B\  \\ \cline{4-6}
atom &  of position &  symmetry &  \ \ \ x\ \ \  & \ \ y\ \  &\  z\  &\\ \hline
La & $4e$ & $4mm$ & 0 & 0 & 0.361 & 1.6 \\ \hline
Sr & $4e$ & $4mm$ & 0 & 0 & 0.361 & 2.0 \\ \hline
Cu & $2a$ & $4/mmm$ & 0 & 0 & 0 & 2.0 \\ \hline
O(I) & $4c$ & $mmm$ & 0 & 0.5 & 0 & 2.5 \\ \hline
O(II) & 4$e$ & $4mm$ & 0 & 0 & 0.183 & 3.5 \\ \hline
\end{tabular}
\label{tab2}
\end{table}

\begin{figure}[]
\includegraphics[width=8cm,angle=0]{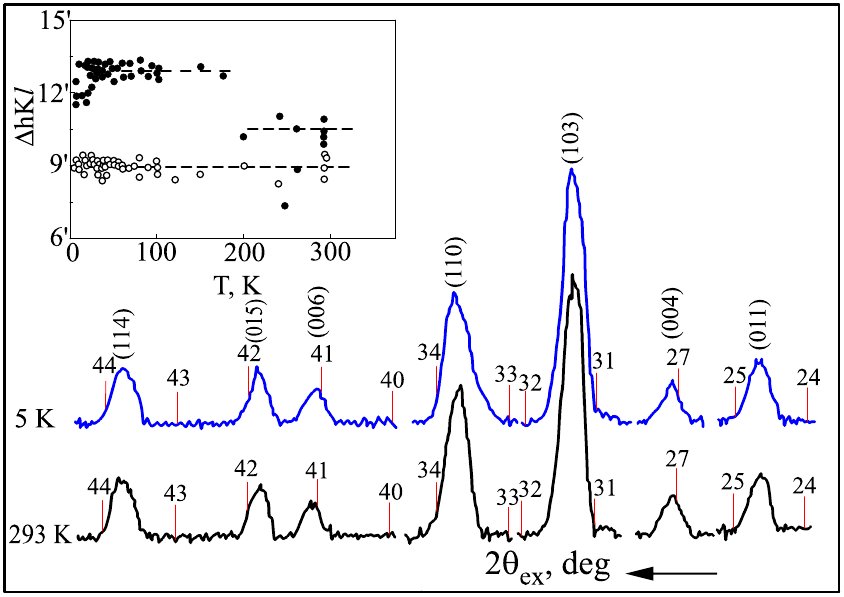}
\caption[]{Typical diffractograms of $\rm La_{2-x}Sr_xCuO_4$ at temperatures of 5~$K$ and 293~$K$. In the inset are shown the temperature dependences of the widths of the (110) x-ray relfection ($\bullet$) and the
103 ($\circ$).}
\label{Fig4}
\end{figure}

During the cooling of samples of a given composition all the way to helium temperatures, no significant changes in the x-ray pattern were detected. On this basis, it can be concluded that the established symmetry of the tetragonal phase of $\rm La_{2-x}Sr_xCuO_4$ is maintained in the temperature interval 4.5-293~$K$ within the limits of accuracy of the implemented x-ray technique. However, at low temperatures ($T < 200\ K$), it was noted that the (110) reflection on the diffractograms has an apparent asymmetry, which is not detected at room temperature (Fig.\ref{Fig4}). Such a change in the diffraction picture can be partly related to the appearance at this angles of a new line.
On behalf
of this proposition, the temperature dependence of the (110) line width, shown in the insert in Fig.\ref{Fig4}, also testifies. At low temperatures, the width of this line is on average 20\% larger than at 293~$K$. However, in the temperature region of the transition to the superconducting state there is a tendency towards a decrease in the (110) line width. The additional line, not corresponding to the spatial group $I4/mmm$ was indeed detected in the photo- x-ray patterns, taken in order to increase the resolution of the reflections in a 360~$mm$ diameter cassette over the course of 1.5-2 hours. We focused on the (110) reflection. It was found in the study of such x-ray patterns that an additional weak line can be seen on the side of small angles near the (110) reflection of the volume-centered tetragonal lattice. Reliable separation and determination of its angular position were hampered by its low intensity and its location on the falling wing of the strong (110) line. The appearance at small angles of this line testifies most probably to change in the symmetry of the tetragonal crystals of $\rm La_{2-x}Sr_xCuO_4$ upon lowering of the temperature. Transition to a different symmetry is a result, apparently, of restructuring in the crystals of the mutual orientation of the $\rm CuO_6$ octahedra or alternating distortions of their equatorial $\rm CuO_4$ squares. At room temperature, stretched
along their own axes, the $\rm CuO_6$ octahedra are oriented within the $\rm La_{2-x}Sr_xCuO_4$ lattice parallel to the fourth order axis (Fig.\ref{Fig3}). Upon lowering of the temperature, periodic deviations arise of small angle of the orientations of the octahedra from the (100) direction or rhombic distortions of the $\rm CuO_4$ squares which alternate from node to node of the lattice. As a consequence of this, the natural symmetry of the octahedra varies and a doubling of the volume of the unit cell occurs as a result of increase of the lattice parameter in the base surface: $a=a_0\sqrt{2}$ where $a_0$ is the parameter of the high-temperature phase. The parameter $c$ is the same for both phases throughout. This leads to a transformation of the volume-centered tetragonal cell to base-centered orthorhombic. It should be noted that such a transition in the basic ceramic $\rm La_2CuO_4$ takes place around 520~$K$. Transformation of the indices of the planes of the reflections during the transition to the new phase corresponds to the following rule: $h' = h + k$, $k' = h - k$, $l'=l$, $h'$, $k'$ and $l'$ are the indices of the planes in the low-temperature phase).

This study of the lattice parameters and estimation of the thermal expansion has shown that there is a substantial anisotropy of the coefficients of thermal expansion for crystals of $\rm La_{2-x}Sr_xCuO_4$, as for other layered structures. Thus, the average values of the coefficients of linear thermal expansion in the base plane and along the direction perpendicular to it in the temperature interval 4.5-293~$K$ are equal to $\bar{\alpha}_a= 0.5\cdot 10^{-5}K^{-1}$ and $\bar{\alpha}_c= 1.7\cdot 10^{-5}K^{-1}$, respectively. In agreement with preliminary data, in the temperature region 25-40~$K$ peculiarities in the behavior of the lattice parameters and anomalies of the linear and volume coefficients of thermal expansion were detected.
\begin{figure}[]
\includegraphics[width=8cm,angle=0]{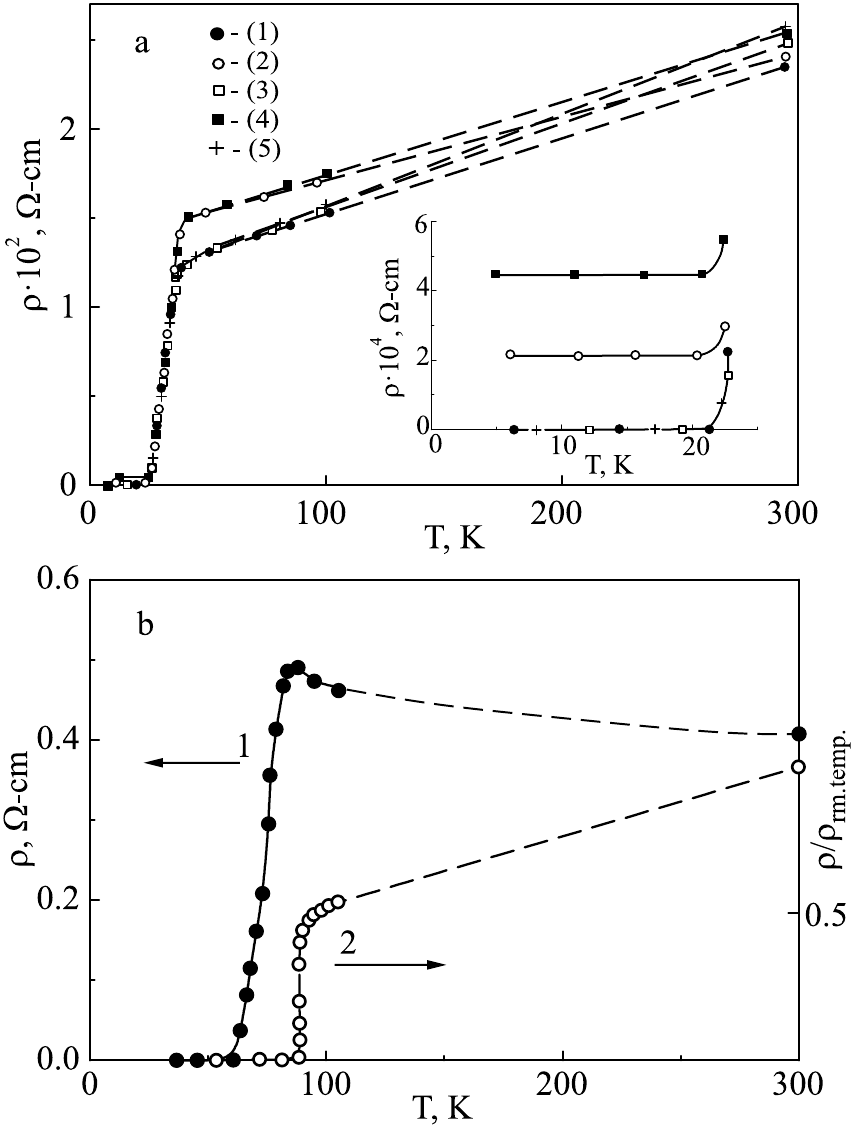}
\caption[]{Temperature dependence of a) electrical resistivity of samples of $\rm La_{1.8}Sr_{0.2}CuO_4$ over the course of 28 days of measurement: (1)- exposure to measurements within 2 days, (2)- 11 days, (3)- 14 days, (4) - 17 days, and (5) - 28 days and b) electrical resistivity of stable compounds of (1) $\rm Y_{1.2}Ba_{0.8}CuO_4$ and 2) $\rm YBa_2Cu_3O_{7-y}$.}
\label{Fig5}
\end{figure}
\begin{figure*}[]
\includegraphics[width=14cm,angle=0]{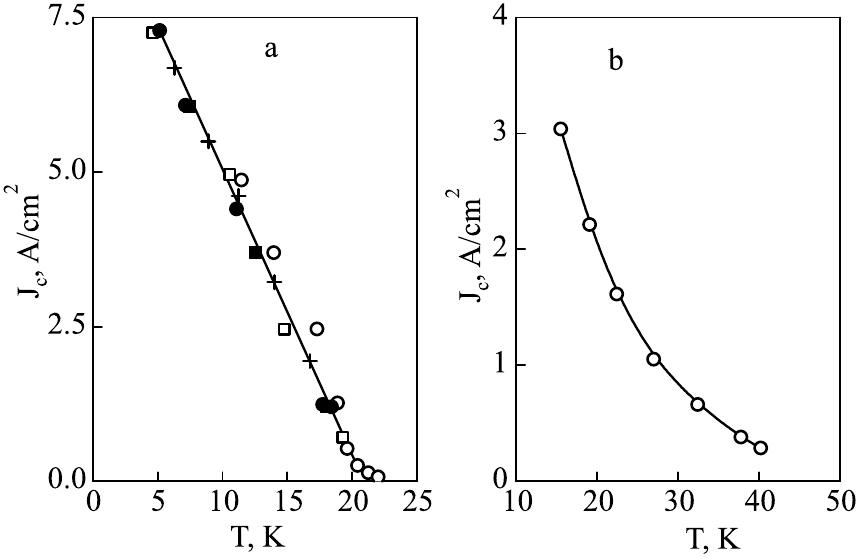}
\caption[]{Temperature dependence of critical current density for a) massive and b) thin samples of $\rm La_{1.8}Sr_{0.2}CuO_4$ (Symbols are the same as in Fig.\ref{Fig5}).}
\label{Fig6}
\end{figure*}
\begin{figure}[]
\includegraphics[width=8cm,angle=0]{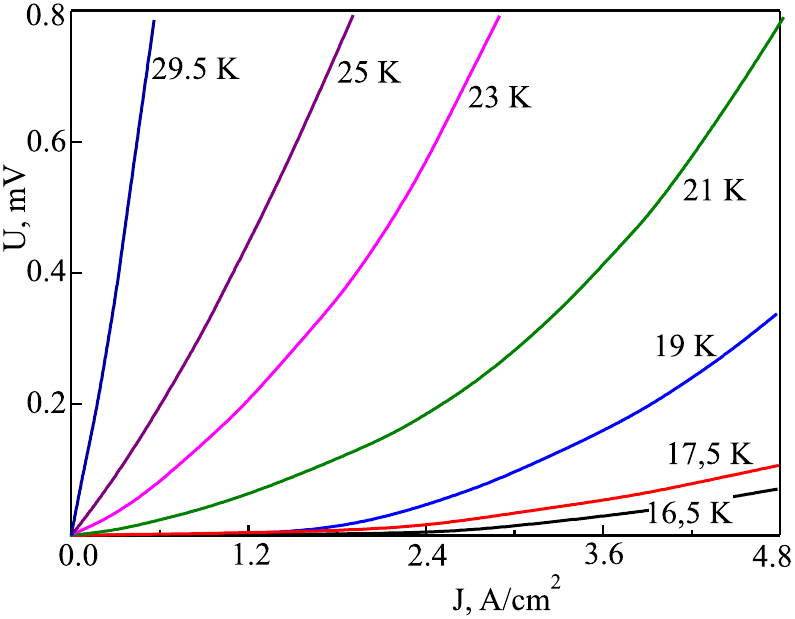}
\caption[]{Temperature dependence of the current-voltage characteristic for samples of $\rm La_{1.8}Sr_{0.2}CuO_4$.}
\label{Fig7}
\end{figure}
\begin{figure}[]
\includegraphics[width=8cm,angle=0]{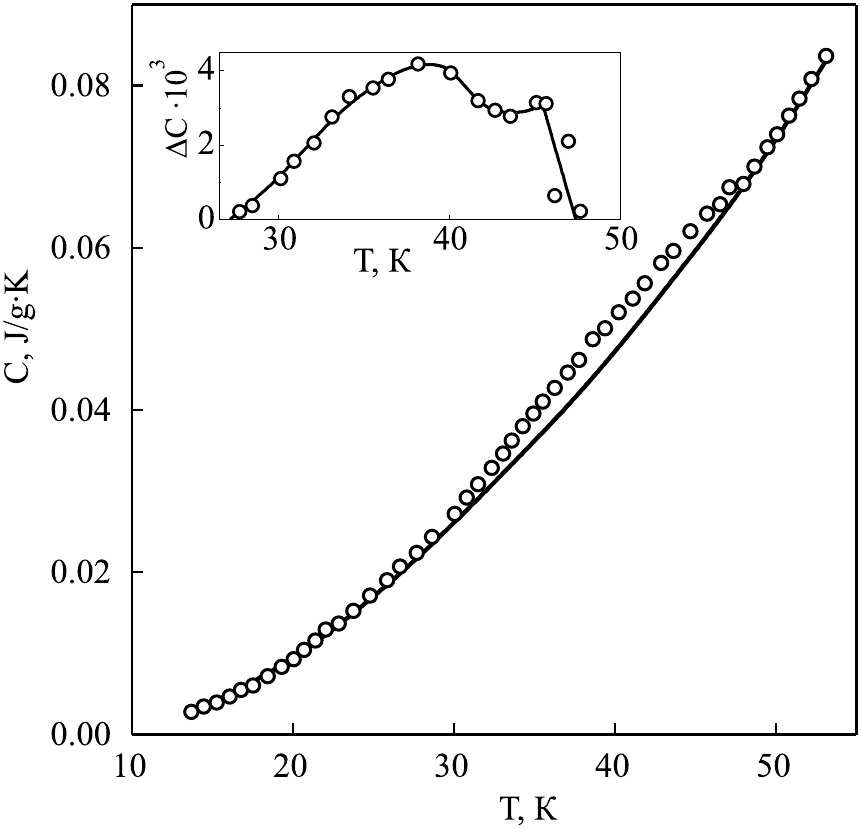}
\caption[]{Temperature dependence of thermal capacity for $\rm La_{1.8}Sr_{0.2}CuO_4$}
\label{Fig8}
\end{figure}
\begin{figure*}[]
\includegraphics[width=15cm,angle=0]{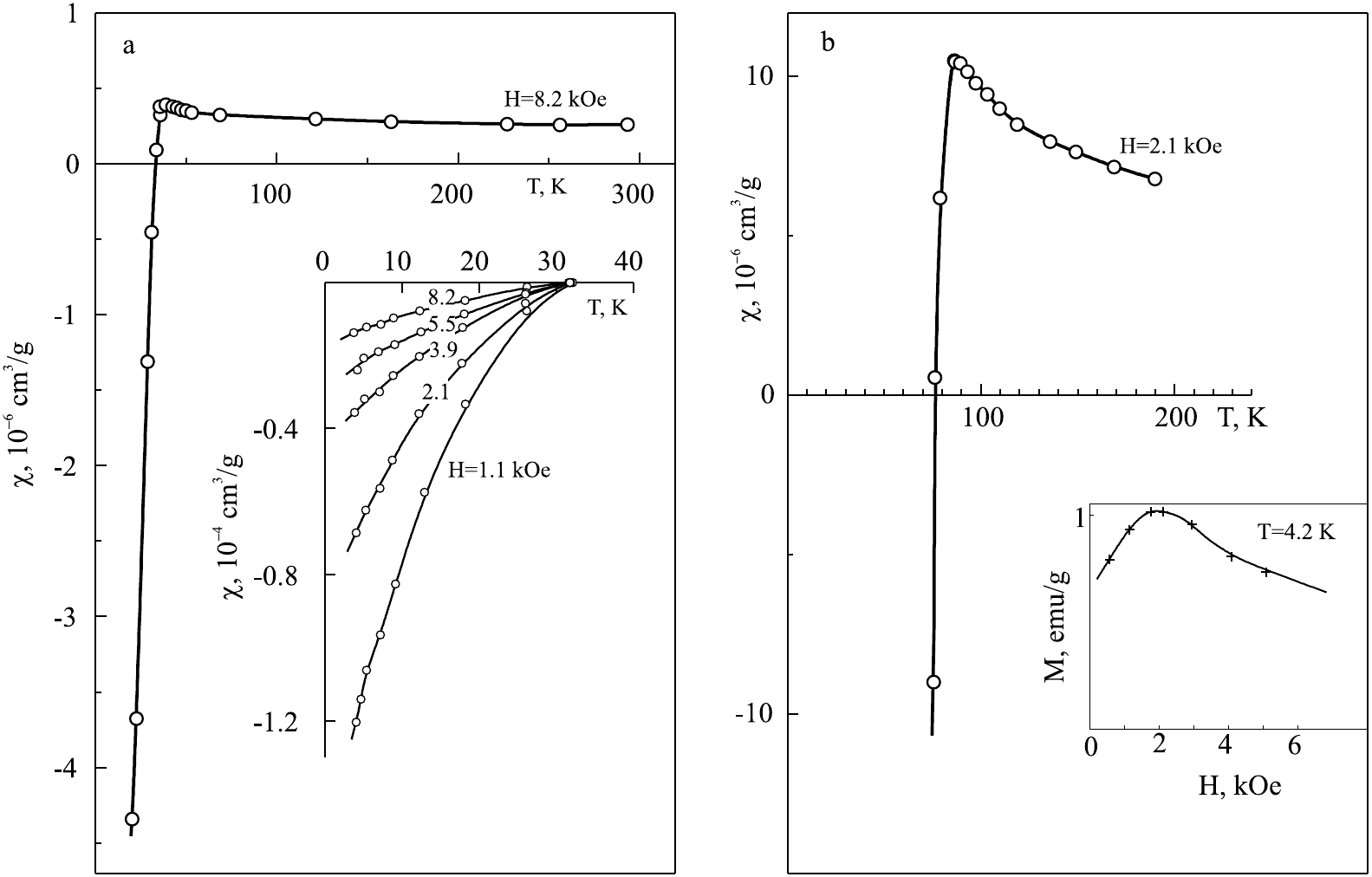}
\caption[]{a) Temperature dependence of the magnetic susceptibility for samples of $\rm La_{1.8}Sr_{0.2}CuO_4$ and b) temperature dependence of the magnetic susceptibility and dependence on field strength $H$ of the magnetic moment $M$ at $T= 4.2\ K$ for samples of Y-Ba-Cu-O.}
\label{Fig9}
\end{figure*}
\section{ELECTRICAL RESISTIVITY OF THE SAMPLES}
We studied the temperature dependence of the electrical resistivity of the superconductors $\rm La_{2-x}Sr_xCuO_4$, $\rm Y_{2-x}Ba_xCuO_4$, and $\rm YBa_2Cu_3O_{7-y}$ by the four-probe method. Degradation of electrical resistance of the samples with time (Fig.\ref{Fig5}) was studied. In the course of 47 days of observation it was established that the superconductivity in a sample of $\rm La_{2-x}Sr_xCuO_4$ at liquid helium temperature disappeared (residual resistance was observed) twice (on the 11th and 17th days); however, afterwards, the sample's superconducting properties were restored.

Electrical resistance measurements at superlow temperatures down to 21~$mK$ were carried out with the goal in mind of investigating the possibility of relapsed superconductivity. The sample was placed in thermal contact with the mixing chamber of a $\rm ^3He-^4He$ dilution refrigerator: with the help of a special lacquer it was glued to the body of the chamber across a thin layer of dielectric. Two resistance thermometers were used to measure temperature: a high-temperature Allen-Bradley thermometer which was placed in thermal contact with the body of the mixing chamber, and a low-temperature SPIR thermometer located within the liquid of the mixing chamber. No appearance of electrical resistance was recorded upon cooling of the sample to 21~$mK$ during which a maximum current density of $2~A/cm^2$ did not lead to destruction of the superconducting state. It may be noted that upon reheating of the sample to room temperature its electrical resistance did not return to its original value; however, it stood only 2\% below it. In a second cooling of the sample down to 77~$K$, its electrical resistance increased to its original value. Such a trend in the electrical resistance is indicative of some kind of mechanical strains in the samples due to thermocycling.
\begin{figure}[]
\includegraphics[width=8cm,angle=0]{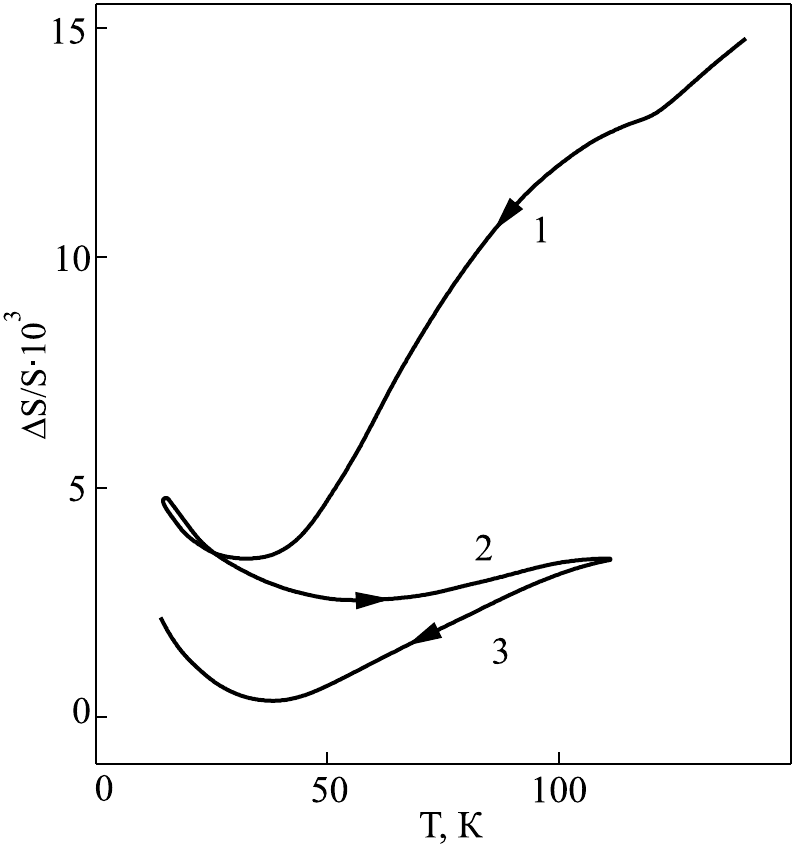}
\caption[]{Temperature dependence of the speed of sound in $\rm La_{1.8}Sr_{0.2}CuO_4$ during thermocycling.}
\label{Fig10}
\end{figure}
\section{CRITICAL SUPERCONDUCTING CURRENT}
The temperature dependence of the critical current density for samples of $\rm La_{1.8}Sr_{0.2}CuO_4$ is shown in Fig.\ref{Fig6}. It may be noted that degradation of the critical current in the samples is not observed.

Measurements of the electrical resistance and critical current of thin samples were carried out via a specially elaborated impulse scheme, which allowed us to keep the influence of thermal effects in the contacts on the superconducting parameters to a minimum. In this technique a current pulse of such a magnitude, that the potential across the sample reaches the threshold value, is passed through the sample. In Fig.\ref{Fig6}b the temperature dependence of the critical current density of thin samples (with cross section $\lesssim 10^{-2}\ cm^2$), measured by such an impulse technique, is shown. The large value of the quantity $|\partial j_c/\partial t|$ for such superconductors should be noted. The current-voltage characteristic of the superconductors is shown in Fig.\ref{Fig7}. For less than critical currents no resistivity in the superconducting state was recorded upon passage of a low-frequency ($\approx 3\ Hz$) variable current of high amplitude.

\section{HEAT CAPACITY OF THE SUPERCONDUCTORS}

Measurement of the heat capacity of the superconductors was carried out by the method of absolute calorimetry over the temperature interval 14-300~$K$. A TSPN- 2V platinum resistance thermometer was used to measure temperature. For example, a sample of $\rm La_{1.8}Sr_{0.2}CuO_4$ with mass greater than 2~$gm$ was placed in a copper container-calorimeter, the mass of which together with the thermometer was around 3~$gm$. The calorimeter was
filled with helium gas and hermetically sealed. The contribution to the heat capacity of the sample to the total heat capacity of the system sample-calorimeter was 30\% at 15~$K$ and 40\% at 50~$K$.

Measurements were made stepping by $T= 0.5-0.8\ K$. Results are shown in Fig.\ref{Fig8}. From the investigated temperature range we can separate out the region 28-47~$K$, ir which there is an anomaly in the heat capacity in the form of a bell-shaped curve. In the insertion in Fig.\ref{Fig8} we show the difference between the experimental values of the heat capacity and a regular curve. The regular curve (solid line in Fig.\ref{Fig8}) was obtained by extrapolating the dependence $C_P(T)$ observed above and below the temperatures of the region under question. The maximum deviation from the regular curve occurs at $T=38\ K$ and stands at 10\%. The anomaly ends in a short step (around 4\%) at a temperature of 42-44~$K$.

\section{MAGNETIC SUSCEPTIBILITY OF THE SUPERCONDUCTORS}

Measurements of the magnetic susceptibility of the superconductors $\rm La_{2-x}Sr_xCuO_4$ and YBaCuO were carried out by the Faraday method in fields of 0.05-0.85~$T$ over the temperature interval 4.2-300~$K$. The investigated sample of the ceramic $\rm La_{1.8}Sr_{0.2}CuO_4$ possesses weak paramagnetism at room temperature ($\chi=0.26\cdot 10^{-6}\ cm^3/g$), which abruptly turns into diamagnetism at a temperature around 35~$K$. The depth of the diamagnetic collapse grows with decrease of the intensity of the magnetic field $H$ (Fig.\ref{Fig9}a). No noticeable indications of degradation of the superconducting characteristics of the material were detected via its magnetic properties over a period of seven days.
\section{PECULIARITIES OF SOUND TRANSMISSION IN SUPERCONDUCTORS}

The temperature dependence of the speed of a longitudinal sound wave with frequency 54.2~$MHz$ in a sample of $\rm La_{1.8}Sr_{0.2}CuO_4$ was measured in the temperature interval 10-130~$K$ (Fig.\ref{Fig10}). The value of the absolute speed of sound was found from acoustical delay measurements to be $(3.5\pm 0.15)\cdot 10^3\ m/sec$. Measurements of the variation of the speed of sound were made by comparing the phases of the last signal through the sample and the comparison signal coherent with it, transmitted to the receiver through an electromagnetic delay line. In order to distinguish the auxiliary signal from the probing signal, we used an acoustic delay line of monocrystalline germanium with a known temperature dependence of its speed of sound. Results of our measurements show a tendency towards a softening of the elasticity modulus in the temperature interval 30-50~$K$. This is indicative of the fact that the system $\rm La_{2-x}Sr_xCuO_4$ undergoes a structural transition or something similar to it in this temperature interval.
\section{POINT-CONTACT SPECTROSCOPY OF THE SUPERCONDUCTING CERAMIC $\rm La_{1.8}Sr_{0.2}CuO_4$}
\begin{figure}[]
\includegraphics[width=7cm,angle=0]{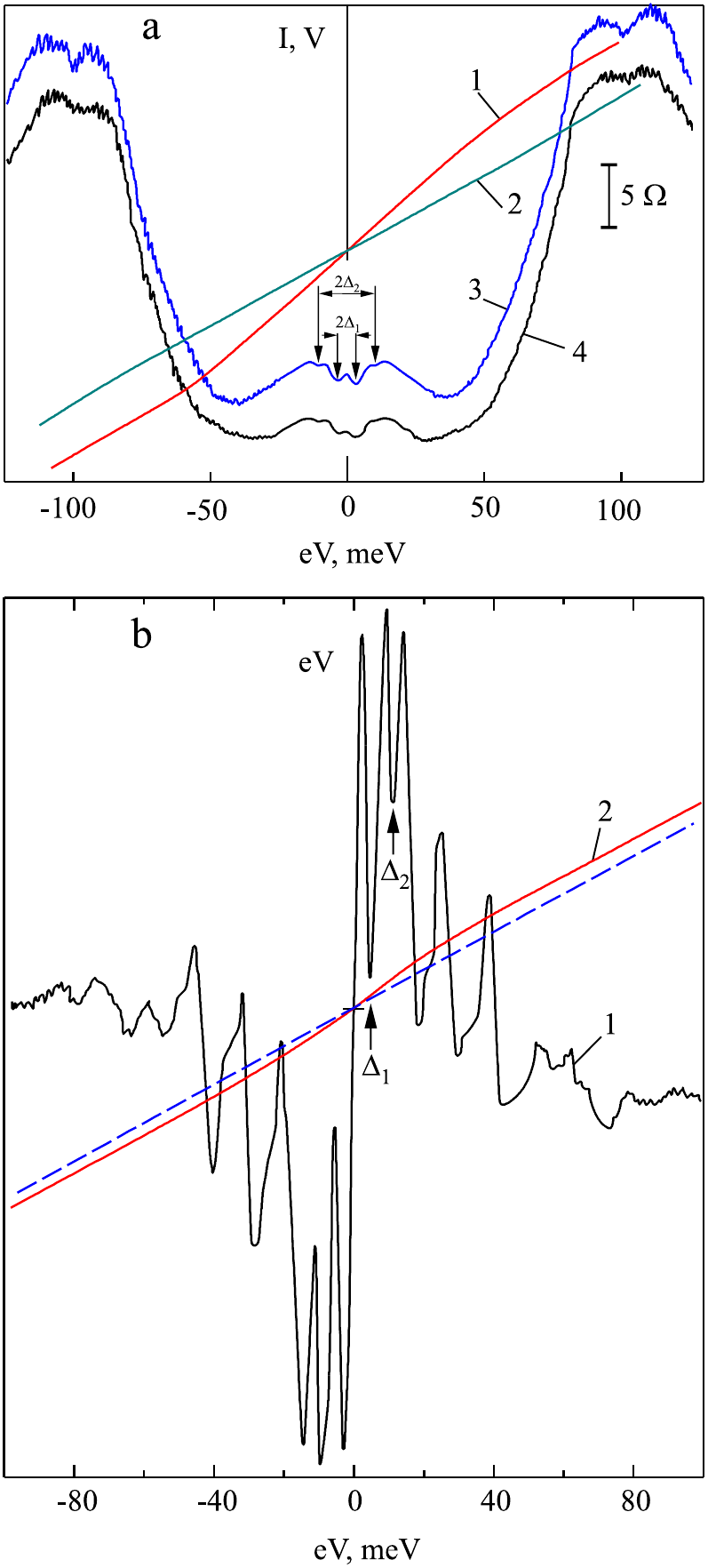}
\caption[]{a) IVC's and their first derivatives for the PC $\rm La_{1.8}Sr_{0.2}CuO_4-Cu$ ($R_N\simeq 120\ \Omega$, $T_{c1}= 34\ K$): curves 1 and 2 are the IVC's at $T=4.2$ and 36~$K$, respectively; curves 3 and 4 are $V_1(eV)$ for $T= 4.2\ K$ for $H=0$ and $H\simeq 15\ kOe$, respectively); b) 1) PC spectrum ($V_2\sim d^2V/dI^2$) and 2) IVC of the ceramic-copper contact ($R_N\simeq 45\ \Omega$) at $T=4.2\ K$ (dashed curve is the assumed IVC in the N-state).}
\label{Fig11}
\end{figure}
The determination of the energy gap and the characteristic frequencies of the lattice vibrations in the material under study was made with the help of the method of point-contact (PC) spectroscopy \cite{6}. Clamping ceramic-copper heterocontacts served as the object of study.

A typical current-voltage characteristic (IVC) and its first derivative ($V_1\sim dV/dI$) of one of the contacts on which gap peculiarities have been detected is displayed in Fig.\ref{Fig11}a (the gap peculiarities in $V_1$ are marked by arrows). (A sharp difference of the Fermi parameters $p_F$ and $v_F$ at the electrodes forming the point-contact, and in addition the presence in the region under consideration of admixtures and structural defects gives the IVC a tunnellike character, where $\Delta$ appears in the form of a maximum on $V_1(eV)$ \cite{7}.

The relative smallness of the excess current $I_{exc}$ formed in the processes of Andreev reflection of quasi-particles from the N-S boundary, and also the low intensity of the gap peculiarities indicate that the PC contact region is made up predominantly of the nonsuperconducting phase. It is possible to explain the appearance of wide minima on $V_1(eV)$ within the framework \cite{8} of the cluster model of the PC (in Fig.\ref{Fig11}a, near 40~$meV$) as a result of superposition of semiconducting and metallic types of conduction (taking into account $I_{exc}$), belonging to the N and S phases, respectively. In a similar way we can also explain the appearance of the wide maximum near 100~$meV$ (curve 3 in Fig.\ref{Fig11}a). This conclusion is confirmed by the circumstance that for a variety of contacts
the location of the indicated peculiarities (in distinction with the gap peculiarities) fluctuates over wide limits.

Thus, the wide minimum near 40~$meV$ cannot be identified with the gap peculiarity; the following fact is indicative of this, namely, that the degree of influence of a magnetic field on the minimum under discussion is significantly higher than on the gap peculiarity (curve 4 in Fig.\ref{Fig11}a).

In the majority of contacts we have observed two gaps each, the maximum values of which were approximately 5 and 12~$meV$, from which we obtain for the ratio $2\Delta/kT_c$ the values 3.3 and 8.0, respectively. In those cases in which only one gap is observed, the corresponding magnitudes had intermediate values. We note that the values of $\Delta$ and $2\Delta/kT_c$ obtained here are close to the values obtained in a previous work \cite{9}, in which a ceramic of similar composition was studied, which had a significantly greater relative volume of superconducting phase.

The PC spectra (the second derivative of the CVC: $V_2\sim d^2V/dI^2$), reflecting the energy distributions of the characteristic quasiparticle excitations in the ceramic under study, were measured in contacts with relatively low $I_{exc}$), for which the intensity of the above-mentioned parasitic peculiarities are negligibly small.

A typical PC spectrum is displayed in Fig.\ref{Fig11}b. The peaks located in the energy region 15-80~$meV$ can be related to phonon modes, as follows from the calculation of Weber \cite{10}. Of special interest are two maxima which are located at significantly higher energies (close to 120 and 180~$meV$), extremely far from the calculated values of Weber. It may be assumed that these maxima result from surface plasmons in $\rm La_{1.8}Sr_{0.2}CuO_4$, through which strong Cooper pairs may be produced in this material \cite{11}.
\section{CONCLUSION}
Structural investigations, the study of electrical resistivity, magnetic susceptibility, heat capacity, acoustical properties, investigation of the point-contact spectra indicate the presence of peculiarities near the superconducting transition temperature.

\end{document}